\documentclass[fleqn,usenatbib]{mnras}
\usepackage{layouts} 

\usepackage{newtxtext,newtxmath}
\usepackage[T1]{fontenc}
\usepackage{xcolor}
\usepackage[utf8]{inputenc}
\usepackage{CJKutf8}
\usepackage{subcaption}
\usepackage{bm}
\usepackage{multirow}
\defcitealias{bogdanov_constraining_2019}{B19}
\defcitealias{choudhury_exploring_2024}{C24}
\defcitealias{pihajoki_general_2018}{P18}

\DeclareRobustCommand{\VAN}[3]{#2}
\let\VANthebibliography\thebibliography
\def\thebibliography{\DeclareRobustCommand{\VAN}[3]{##3}\VANthebibliography}

\author[W. Zhang \& W. Yu]{
Wenda Zhang,$^{1}$\thanks{wdzhang@nao.cas.cn}, Wenfei Yu$^{2}$\\
$^1$National Astronomical Observatories, Chinese Academy of Sciences, A20 Datun Road, Beijing 100101, China \\
$^2$Shanghai Astronomical Observatory, Chinese Academy of Sciences, 80 Nandan Road, Shanghai 200030, China
}

\usepackage{graphicx}	

\newcommand\monk{\textsc{Monk}}
\newcommand\codename{\textsc{Monk-NS}}

\graphicspath{{./figures/}}
\def\sd1width{0.49}
\def\os1width{0.49}
\def\c24width{0.49}
\def\burstwidth{0.49}

\title[Neutron star pulsation profile]{Modeling surface radiation of rotating neutron stars with \codename{}}

\date{Accepted XXX. Received YYY; in original form ZZZ}

\pubyear{2025}

\begin{document}
\label{firstpage}
\pagerange{\pageref{firstpage}--\pageref{lastpage}}
\maketitle

\begin{abstract}
Neutron stars serve as unique laboratories for studying ultra-dense nuclear matter. The equation of state of neutron star matter can be effectively constrained by their masses and radii. Particular attention has been paid to rapidly rotating neutron stars, where strong relativistic effects leave imprints on their electromagnetic emission. To model the emission of rotating neutron stars in more realistic situations, especially when their surface emission is further re-processed by a scattering medium, we develop \codename{}, a customized version of the general relativistic Monte-Carlo radiative transfer code \monk{}. We validate the code through a series of benchmarking tests, including computing the energy spectrum, pulse profile, and polarisation of rotating neutron stars, and comparing the results with those of the established codes in the X-ray timing community, yielding consistent outcomes. As an example to demonstrate \codename{}'s capabilities, we apply it to investigate various models proposed to explain the low pulsation amplitude of neutron star low-mass X-ray binaries. Our findings indicate that the dependence of the X-ray polarisation degree on the observer's inclination can serve as a key factor in distinguishing these models. We also find that complex hotspot morphologies yield polarisation properties different from those of circular hotspots.
\end{abstract}



\section{Introduction}

The physical properties, including the interactions and the compositions of ultradense matter, are still highly uncertain \citep[e.g.][for reviews]{hebeler_three-nucleon_2021}. Neutron stars (NSs), the densest known astrophysical objects - with a core density of $\sim 10^{14}~\rm g~cm^{-3}$ - turn out to be ideal astrophysical laboratories for studying the physics of dense matter. While the equation of state (EoS) of neutron stars is not directly observable, it can be constrained through measurements of their mass and radius \citep[e.g.][]{lattimer_neutron_2007,lattimer_nuclear_2012,watts_colloquium_2016,lattimer_neutron_2021}.


For pulsating neutron stars, the mass and radius can be determined by modeling their X-ray pulse profile. This is because the mass and radius of a neutron star significantly influence the magnitudes of several specific relativistic effects, such as the Doppler effect, and general relativistic effects, like the light-bending effect. These effects play a crucial role in shaping the pulsation profile \citep[see, e.g.][]{miller_bounds_1998,braje_light_2000,beloborodov_gravitational_2002,poutanen_nature_2003,poutanen_pulse_2006,viironen_light_2004,lamb_model_2009,loktev_oblate_2020,ahlberg_effects_2024}. This method is particularly effective for rapidly-spinning neutron stars (spin frequency $\nu_{\rm spin} \gtrsim 100~\rm Hz$), where the relativistic effects are more profound.

There are several different classes of neutron stars that show fast (spin period of a few milliseconds) X-ray oscillations. Among them, rotation-powered millisecond pulsars (MSPs) that exhibit thermal X-ray pulsations with temperatures of $\sim 10^6~\rm K$ \citep{zavlin_xmm-newton_2006,zavlin_studying_2007} are particularly well-suited for constraining neutron star mass and radius through pulse profile modeling.
The periodic X-ray emission is believed to originate from hotspots on the neutron star surface heated by the backflow of energetic particles accelerated by the magnetic field \citep[e.g.][]{zhang_full_2000,zhang_x-ray_2003}. Strong constraints (typically $\lesssim 5\%$) have been put on the radius of the neutron stars in MSP systems through pulse profile modeling \citep[e.g.][]{miller_psr_2019,riley_nicer_2019,raaijmakers_nicer_2019,miller_radius_2021,salmi_radius_2022,vinciguerra_updated_2024}, especially after the launch of the \textit{Neutron Star Interior Composition Explorer} \citep[\textit{NICER};][]{gendreau_neutron_2016}.

The pulse profile modeling method has also been successfully applied to accreting millisecond X-ray pulsars (AMXPs) \citep[see e.g.][for reviews]{patruno_accreting_2021, di_salvo_accretion_2022}, first discovered in 1998 by \citet{wijnands_millisecond_1998}. AMXPs are low-mass X-ray binaries (LMXBs) in which the millisecond pulsar accretes material from its low-mass companion, and are believed to be progenitors of rotation-powered MSPs. The accretion flow is channeled by the magnetic field onto the surface of the neutron star, creating hotspots at the magnetic poles and producing coherent pulsations. To date, approximately 21 AMXPs have been discovered \citep{di_salvo_accretion_2022, sanna_maxi_2022}. Several measurements of the neutron star radius and mass have been made for AMXPs \citep[e.g.][]{poutanen_nature_2003, leahy_limits_2008, leahy_constraints_2009, leahy_constraints_2011, morsink_multi-epoch_2011}, although with larger uncertainties (typically on the order of tens of percent). It is important to note that AMXPs constitute only a small fraction of the NSLMXB population, and the reason why most NSLMXBs do not exhibit pulsations remains an open question.

NSLMXBs (not necessarily AMXPs) exhibit oscillations during type-I thermonuclear bursts, known as burst oscillations \citep{strohmayer_363_1997} \citep[see][for a review]{galloway_thermonuclear_2021}. For a given source, the frequencies of burst oscillations measured in different bursts are found to fall within a narrow range, suggesting that the burst oscillation is modulated by the spin frequency. This was later confirmed by the detection of burst oscillations in a few AMXPs \citep{chakrabarty_nuclear-powered_2003}. Type-I bursts are triggered by unstable thermonuclear runaways, occurring when the heating rate from nuclear burning exceeds the radiative cooling rate. Burst oscillations can be observed if either the thermonuclear burning is confined to a small region, creating a hotspot, or if a global mode is excited following the initial ignition \citep{watts_thermonuclear_2012}. In both cases, the mass and radius of the neutron star can be constrained by modeling the pulsation profile of the burst oscillation. Several studies have attempted to do so \citep{nath_bounds_2002, bhattacharyya_constraints_2005}, though only upper limits have been obtained.
It is anticipated that future observations with missions featuring larger collecting areas will provide much tighter constraints \citep{lo_determining_2013, miller_determining_2015}.


The radius and mass of a neutron star can also be determined by modeling the energy spectrum of accreting neutron stars, particularly in quiescent NSLMXBs in globular clusters \citep[e.g.][]{heinke_hydrogen_2006, guver_mass_2010, guillot_neutron_2011, guillot_discovery_2011, servillat_neutron_2012, catuneanu_massradius_2013, lattimer_neutron_2014, bogdanov_neutron_2016, steiner_constraining_2018, shaw_radius_2018, echiburu_spectral_2020} and accreting NSs exhibiting photospheric radius expansion during thermonuclear bursts \citep[e.g.][]{ozel_mass_2009, guver_mass_2010, guver_distance_2010, kusmierek_mass_2011, suleimanov_neutron_2011, zamfir_constraints_2012, ozel_mass_2012, guver_mass_2013}. Additionally, constraints on the radius of neutron stars can be obtained through phase-resolved spectral analysis when milli-Hz quasi-periodic oscillations are observed \citep[e.g.][]{stiele_millihertz_2016}.

In addition to timing and energy spectrum, the phase-averaged and phase-resolved X-ray polarisation of MSPs can provide us with valuable information of the neutron stars. \citet{viironen_light_2004} studied the X-ray polarisation and pulsation profiles of AMXPs and bursting NSLMXBs using a Thomson formalism and assuming the Schwarzschild plus Doppler approximation, and found the polarisation angle to change substantially compared to the results assuming the spherical Schwarzschild metric. \citet{ahlberg_effects_2024} studied the effects of scatterings on the pulse profile of AMXPs. \citet{loktev_oblate_2020} investigated the X-ray polarisation assuming the oblate Schwarzschild approximation, accounting for the deformation of the neutron star due to rapid rotation of the neutron star.

While analytical solutions can be obtained for simple problems, numerical methods have to be employed for more realistic and/or complicated configurations, especially when the surface radiation is further reprocessed by a scattering medium.
To model the radiation of rotating neutron stars with more realistic and/or complicated configuration, we develop \codename{}, a Monte-Carlo radiative transfer code customized for neutron stars. The Monte-Carlo nature of the code allows for great flexibility, enabling us to deal with emitting regions with arbitrary geometrical and/or physical properties. In this paper, we give an introduction to the code and present a simple application of the code. More complicated applications of the code will be presented in a series of forthcoming papers.

The paper is organized as follows: In Sec.~\ref{sec:method}, we introduce the methods employed by \codename{}. In Sec.~\ref{sec:validation}, we validate \codename{} by comparing its results with those from the established codes used in the NICER collaboration, demonstrating a high level of consistency. In Sec.~\ref{sec:application}, we apply \codename{} to investigate several scenarios that could explain why most NSLMXBs do not exhibit pulsation. Finally, the results are summarized in Sec.~\ref{sec:summary}.


\section{Method}
\label{sec:method}

\monk{} is a Monte Carlo radiative transfer code that incorporates all general relativistic effects \citep{zhang_constraining_2019}. The code samples photons from various emission processes and traces their paths along null geodesics in curved spacetimes. \monk{} is also capable of handling a range of opacities, including Compton scattering, bremsstrahlung self-absorption, and synchrotron self-absorption. In comparison to the version introduced in the original paper, we have added routines to trace the time ($t$) and azimuthal angle ($\phi$) components of the photon’s four-position. This enhancement allows us to perform time-dependent and/or axially asymmetric simulations \citep[for details, see][]{zhang_theoretical_2023}.

In this paper, we present \codename{}, a numerical code based on \textsc{Monk}, customized to model the radiation from rotating neutron stars.

\subsection{Spacetime}


In \codename{}, we employ two different metrics to approximate the spacetime exterior to the surface of the neutron star: the Kerr metric, and the Hartle-Thorne metric that describes the exterior spacetime of a slowly rotating, stationary, and axially symmetric body. In the latter case, we utilize the Glampedakis \& Babak variant \citep[hereafter GB06 metric;][]{glampedakis_mapping_2006} of the Hartle-Thorne metric. In addition to the specific angular momentum $a$, the GB06 metric has another parameter $\epsilon$ that denotes the deviation from the Kerr metric in the quadrupole moment, such that
\begin{equation}
q \equiv \frac{\mathcal{Q}}{M^3} = -a^2 (1 + \epsilon),
\end{equation}
where $M$ and $\mathcal{Q}$ are the mass and quadrupole moment of the body, respectively.

The wave and polarisation vectors in the co-moving frame $\bm{k}^{(a)},\bm{f}^{(a)}$ can then be transformed into the coordinate frame by
\begin{align}
 \bm{k}^\mu = \boldsymbol{e}^\mu_{(a)} \bm{k}^{(a)},\\
 \bm{f}^\mu = \boldsymbol{e}^\mu_{(a)} \bm{f}^{(a)}.
\end{align}
The energy of a photon measured by an observer at infinity is
\begin{equation}
E_{\rm loc} = -\bm{k}_\mu \bm{U}^\mu E_\infty,
\end{equation}
where $E_{\rm loc}$ is the energy measured by a co-moving observer, and $\bm{U}^\mu$ is the four velocity of the surface fluid. We normalize $\bm{k}^\mu$ in such a way that $k_t = -1$. \codename{} supports two different methods of raytracing. In the first method, we utilise the Hamilton-Jacobi equation to raytrace the coordinates while the wave and polarisation vectors are solved when necessary by making use of the constants of motion \citep{zhang_constraining_2019,zhang_theoretical_2023}. In the second method, both the coordinates and the wave and polarisation of the photons are transported along null geodesic by integrating the geodesic equation using a fourth-order Runge-Kutta method. Depending on the specific problem, we choose whichever method is more convenient.



\subsection{Shape}
 In addition to assuming a simple spherical shape for the neutron star, \codename{} is also capable of modeling the emission from an oblate NS, which is more realistic for rapidly rotating systems. The oblateness affects the pulse profile, with the effect being more pronounced for rapidly rotating NSs \citep[e.g.][]{cadeau_light_2007,morsink_oblate_2007,psaltis_pulse_2014,miller_determining_2015}. \citet{morsink_oblate_2007} found that the oblateness is largely insensitive to the equation of state and proposed a universal relation between the oblateness and the mass, radius, and spin of the NS. For the oblate case, we use the prescription provided by \citet{morsink_oblate_2007} to calculate the shape and surface gravity of oblate neutron stars.


\subsection{Numerical Schemes}
 In \codename{}, we employ two different numerical schemes: the observer-to-emission (o2e) scheme and the Monte Carlo (mc) scheme. While the o2e method is fast and straightforward, the mc scheme is more versatile, capable of handling more complex scenarios, particularly when surface radiation is reprocessed by a scattering medium before reaching the observer.


\subsubsection{The o2e Scheme}
 In this scheme, we set up an image plane with coordinates $\alpha,\beta$ at a large distance from the neutron star, at a given inclination. We generate synthetic images by dividing the image plane into multiple pixels. For each pixel, we raytrace the photon backward in time until it reaches the surface of the neutron star. The flux density observed by the observer is given by
\begin{equation}
 F_\nu = \sum I_\nu(\vartheta, \nu/g)g^3 d\alpha d\beta / D^2,
\end{equation}
where $I_\nu(\vartheta,\nu)$ is the specific intensity of the surface radiation as a function of $\nu$ and the polar emission angle $\vartheta$, $D$ is the distance to the neutron star, and $g\equiv -1/\mathbf{k_\mu} \mathbf{U^\mu}$ is the redshift factor. The summation is performed over all pixels that reach the hotspot.

 To speed up computation, we employ an adaptive mesh refinement method based on the quadtree algorithm, following \citet{davelaar_self-lensing_2022-1}. This method begins with a coarse grid. For each pixel, we compute the flux and evaluate the relative flux differences between the pixel and its four neighbors, denoted as $s_{\rm kl}$. If the mean relative difference exceeds a critical threshold, the pixel is subdivided into four smaller pixels. This refinement process continues recursively until $s_{\rm kl}$ meets the convergence criterion for all pixels or the maximum predefined recursion level is reached.

\subsubsection{The mc Scheme}
 In this scheme, the surface emission of the neutron star is modeled by sampling "superphotons." Each superphoton represents a packet of identical photons and is characterized by $E_\infty$, the linear polarisation degree $\delta$, $\bm{k}^\mu$, $\bm{f}^\mu$, and statistical weight $w$, which corresponds to the number of photons emitted per unit time as measured by a distant observer.


 We begin by dividing the emitting region on the neutron star's surface into small surface elements. For each surface element, we discretize the photon momentum space, compute the wave and polarisation vectors in the co-moving frame, and then transform these vectors into the coordinate frame. The statistical weight of a superphoton is given by \citep{schnittman_monte_2013}:
\begin{equation}
\label{eq:weight}
w = \frac{dA d\Omega {\rm cos}\vartheta} {U^t N_{\rm sp}}\int_0^\infty \frac{I_\nu(\vartheta,\nu)}{h\nu} d\nu,
\end{equation}
where $d\Omega$ is the solid angle subtended by the pixel in photon momentum space, $U^t$ is the time-component of the surface fluid's four-velocity, $N_{\rm sp}$ is the number of superphotons sampled from this pixel, $h$ is Planck's constant, and the proper area of a surface element
\begin{equation}
dA = \gamma R^2({\theta}) {\rm sin}\theta d\theta d\phi.
\end{equation}
Here $\gamma$ is the Lorentz factor of the surface element as seen by a zero angular momentum observer, $\theta$ is the colatitude of the surface element, and $d\theta$ and $d\phi$ are the angular sizes of the surface element in the polar and azimuthal directions, respectively. $R(\theta)$ is the radius of the neutron star as a function of colatitude. For a spherical neutron star, $R(\theta)$ always equals the equatorial circumferential radius $R_{\rm eq}$, but this is no longer true for an oblate neutron star. If the surface of the neutron star emits blackbody radiation such that the specific intensity follows the Planck function, we can derive the statistical weight of one superphoton:
\begin{equation}
w = \frac{4\zeta(3) f_{\rm limb} dA d\Omega {\rm cos}\vartheta \ (k_{\rm B}T_{\rm NS})^3}{U^t h^3 c^2 N_{\rm sp}},
\end{equation}
where $\zeta(3)$ is Ap\'ery's constant, $f_{\rm limb}$ is the limb-darkening factor, $k_{\rm B}$ is the Boltzmann constant, and $T_{\rm NS}$ is the neutron star surface temperature in the local frame.

For each superphoton, we first determine whether it is emitted outward or inward from the neutron star surface. For a spherical NS, this is straightforward by checking the sign of $k^r$. For an oblate NS, we propagate the superphoton by a small step and check whether it is inside the NS surface after propagation. If the superphoton is emitted outward, we parallel transport the four-position $\bm{x}^\mu$, $\bm{k}^\mu$, $\bm{f}^\mu$ to infinity. Meanwhile, $E_\infty$, $\delta$, and $w$ remain constant. This procedure is repeated for all surface elements within the emitting region and for all pixels in photon momentum space. If the superphoton is reprocessed by a scattering medium before reaching the observer, we take into account the Compton scattering using the recipes that are described in detail in \citet{zhang_constraining_2019}.

 Finally, we collect all superphotons arriving at infinity, and store the following information of the photons: $w$, $E_{\infty}$, $\bm{x}^\mu$, $\bm{k}^\mu$, $\bm{f}^\mu$. These data can be used to compute the energy spectrum, polarisation, and lightcurves of an observer at infinity at any inclination while detailed procedures can be found in \citet{zhang_constraining_2019,zhang_theoretical_2023}.

%

\section{Code validation}
\label{sec:validation}
In this section, we conduct several tests and compare the results with previous studies to validate \codename{}. For each test case, we present the outcomes of both the o2e and mc schemes. We will demonstrate that the results obtained with the o2e scheme agree with the literature within 1\%. Although the mc results exhibit larger fluctuations, they gradually converge to the o2e results as the number of superphotons increases.

\subsection{Energy spectrum}
\label{sec:spectrum}
\begin{figure}
\includegraphics[width=\columnwidth]{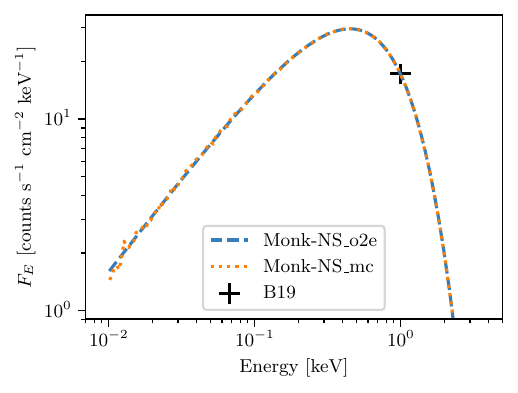}
\caption{The energy spectra of a thermally-emitting, non-rotating neutron star, with $M_{\rm NS}=1.4~\rm M_\odot$, $R_{\rm eq}=12~\rm km$, $T_{\rm NS}=0.35~\rm keV$, and $D=200~\rm pc$. The results obtained with the mc and o2e schemes are plotted in different colors and linestyles, as indicated in the plot. The flux density at $1~\rm keV$ calculated by \citetalias{bogdanov_constraining_2019} is $17.2279~\rm counts~s^{-1}~cm^{-2}~keV^{-1}$, which is marked by a black cross.\label{fig:enspec}}
\end{figure}

We compute the energy spectrum of a uniformly-emitting, non-rotating neutron star, using parameters identical to those specified in Sec. C.2 of \citet{bogdanov_constraining_2019} (hereafter \citetalias{bogdanov_constraining_2019}): the neutron star has a mass of $M_{\rm NS} = 1.4~\rm M_\odot$, a radius of $R_{\rm eq}=12~\rm km$, and is located at a distance of $D=200~\rm pc$. The local emission of the neutron star is assumed to be isotropic and thermal, with a blackbody temperature of $T_{\rm NS}=0.35~\rm keV$. While \citetalias{bogdanov_constraining_2019} did not provide the full spectrum, they gave the flux density at $1~\rm keV$, which is $17.2279~\rm counts~s^{-1}~cm^{-2}~keV^{-1}$. The flux obtained with the o2e scheme is $17.2276~\rm counts~s^{-1}~cm^{-2}~keV^{-1}$, in excellent agreement with \citetalias{bogdanov_constraining_2019}. We also present the energy spectra computed with both schemes in Fig.~\ref{fig:enspec}, with the \citetalias{bogdanov_constraining_2019} result indicated by a black cross. It is clear that the results from the two schemes are consistent, although the mc scheme shows larger fluctuations, as expected.

\subsection{Pulse profile}
\subsubsection{Spherical NSs}
\begin{figure*}
\begin{center}
\begin{subfigure}{\sd1width\textwidth}
 \includegraphics{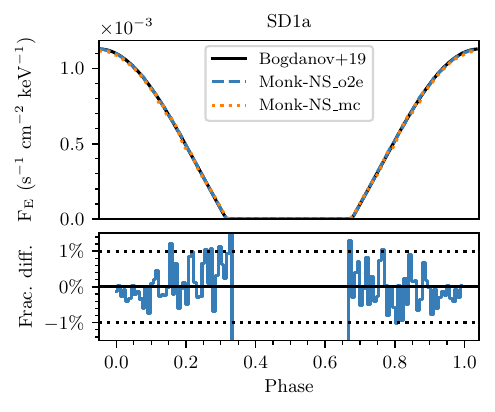}
\end{subfigure}
\begin{subfigure}{\sd1width\textwidth}
 \includegraphics{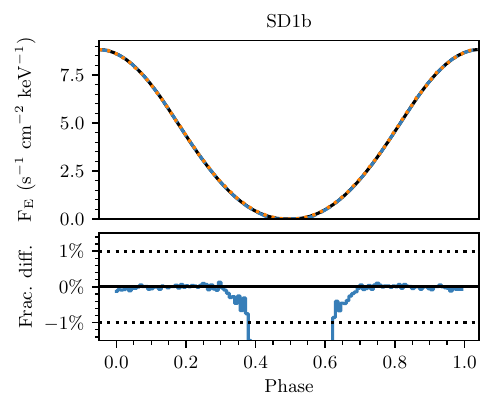}
\end{subfigure}
\begin{subfigure}{\sd1width\textwidth}
 \includegraphics{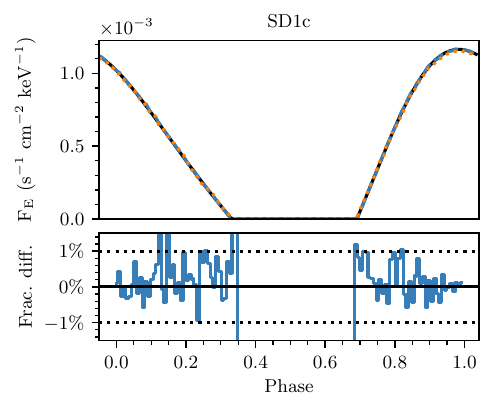}
\end{subfigure}
\begin{subfigure}{\sd1width\textwidth}
 \includegraphics{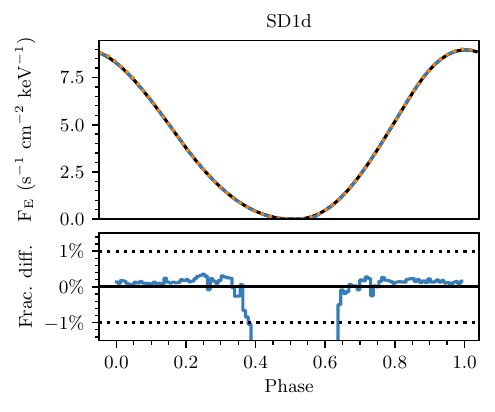}
\end{subfigure}
\begin{subfigure}{\sd1width\textwidth}
 \includegraphics{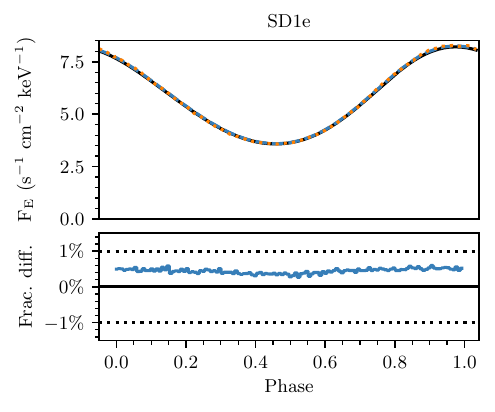}
\end{subfigure}
\begin{subfigure}{\sd1width\textwidth}
 \includegraphics{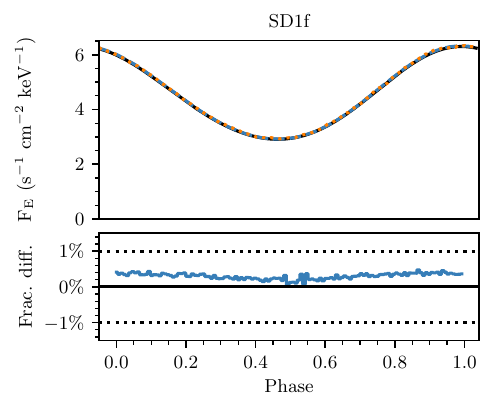}
\end{subfigure}
\end{center}
\caption{A comparison of the pulse profile predictions between \codename{} (dashed blue and dotted orange for the o2e and mc schemes, respectively) and \citetalias{bogdanov_constraining_2019} (black). The parameters used in tests SD1a through SD1f can be found in Table~\ref{tab:sd1}.\label{fig:sd1}}
\end{figure*}

\citetalias{bogdanov_constraining_2019} performed six verification tests, designated SD1a--SD1f, in which the authors computed the pulse profiles of MSPs with various parameters under the Schwarzschild+Doppler (S+D) assumption. Under the S+D assumption, the shape of the neutron star is spherical, and the exterior spacetime is Schwarzschild. The values of the parameters are tabulated in Table~\ref{tab:sd1}. In these tests, a single circular hotspot is assumed. To validate \codename{}, we compute the pulse profiles using \codename{} assuming a spherical NS with identical parameters to those in Schwarzschild spacetime. In Fig.~\ref{fig:sd1}, we present the results obtained with both schemes of \codename{} and compare them with \citetalias{bogdanov_constraining_2019}. For each case, we also compute the fractional difference, and present it in the lower panel of each subplot in Fig.~\ref{fig:sd1}. In all cases, the fractional difference between the lightcurves of the o2e scheme and those of \citetalias{bogdanov_constraining_2019} is smaller than 1\% except for phase bins where the flux is close or equal to zero. The same behavior is seen in the comparison of the pulse profiles produced by different codes in \citetalias{bogdanov_constraining_2019}. Similar to Sec.~\ref{sec:spectrum}, the mc lightcurves agree with both o2e and \citetalias{bogdanov_constraining_2019} results, albeit with larger fluctuations.

\begin{table*}
\begin{center}
\caption{The values of parameters for tests SD1a--SD1f, OS1--OS1j, and SD2a--SD2d. The values are identical with corresponding tests of \citetalias{bogdanov_constraining_2019}.\label{tab:sd1}}
\begin{tabular}{|lllllllll|}
\hline
Test & $M_{\rm NS}$ & $R_{\rm eq}$ & $T_{\rm NS}$ & $\theta_0$ & $R_0$ & $i_{\rm obs}$ & $\nu_{\rm spin}$ & $f_{\rm limb}$\\
& $(M_{\odot})$ &(km)& (keV) & (deg)&(rad)&(deg)&(Hz)\\
\hline
\multicolumn{9}{c}{SD1 tests} \\
SD1a &1.4&12 & 0.35 & 90 & 0.01 & 90 & 1 & 1\\
SD1b &1.4&12&  0.35 &90 & 1.0 & 90 & 1 & 1 \\
SD1c &1.4&12&  0.35 &90 & 0.01 & 90 & 200 & 1 \\
SD1d &1.4&12&  0.35 &90 & 1.0 & 90 & 200 & 1\\
SD1e &1.4&12&  0.35 &60 & 1.0 & 30 & 400 & 1\\
SD1f &1.4&12&  0.35 &20 & 1.0 & 80 & 400 & 1\\
\hline
\multicolumn{9}{c}{SD2 tests} \\
SD2a&1.4&12&  0.35 &90 & 0.01 & 90 & 1 & 1\\
SD2b&1.4&12&  0.35 &90 & 1.0 & 90 & 1& 1\\
SD2c&1.4&12&  0.35 &90 & 0.01 & 90 & 400 & 1\\
SD2d&1.4&12&  0.35 &90 & 1.0 & 90 & 400 & 1\\
\hline
\multicolumn{9}{c}{OS1 tests} \\
OS1a&1.4&12&  0.35 &90 & 0.01 & 90 & 600 & 1 \\
OS1b&1.4&12&  0.35 &90 & 1.0 & 90 & 600 & 1 \\
OS1c&1.4&12&  0.35 &90 & 0.01 & 90 & 200 & 1 \\
OS1d&1.4&12 &  0.35 &90 & 1.0 & 90 & 1 & 1 \\
OS1e&1.4&12&  0.35 &60 & 1.0 & 30 & 600 & 1 \\
OS1f&1.4&12&  0.35 &20 & 1.0 & 80 & 600 & 1 \\
OS1g&1.4&12&  0.35 &60 & 1.0 & 30 & 600 & ${\rm cos}^2\vartheta$\\
OS1h&1.4&12&  0.35 &60 & 1.0 & 30 & 600 & ${\rm sin}^2\vartheta$\\
OS1i&1.4&12&  0.35 &20 & 1.0 & 80 & 600 & ${\rm cos}^2\vartheta$\\
OS1j&1.4&12&  0.35 &20 & 1.0 & 80 & 600 & ${\rm sin}^2\vartheta$\\
OS1k& 1.4&12 & 0.35 & 90 & 0.01 & 90 & 400 & 1 \\
OS1l& 1.4&12 & 0.35 & 90 & 1 & 90 & 400 & 1 \\
\hline
\multicolumn{9}{l}{$\theta_0$: the centroid colatitude of the hotspot;} \\
\multicolumn{9}{l}{$R_0$: the angular radius of the hotspot;} \\
\multicolumn{9}{l}{$i_{\rm obs}$: the observer's inclination;}\\
\multicolumn{9}{l}{$\nu_{\rm spin}$: the NS spin frequency in Hz.}\\
\end{tabular}
\end{center}
\end{table*}

\subsubsection{Oblate NS}
\begin{figure*}
\centering
\begin{subfigure}{\os1width\textwidth}
 \includegraphics{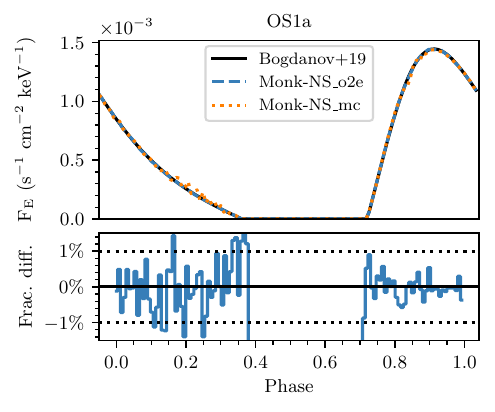}
\end{subfigure}
\begin{subfigure}{\os1width\textwidth}
 \includegraphics{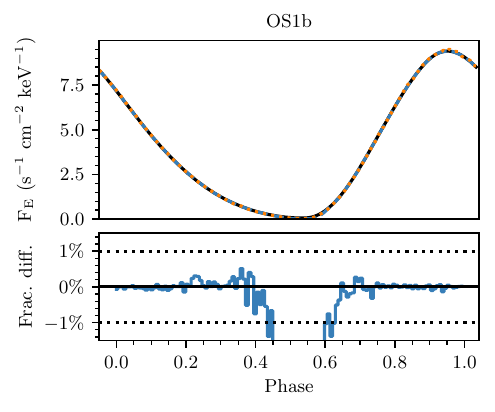}
\end{subfigure}
\begin{subfigure}{\os1width\textwidth}
 \includegraphics{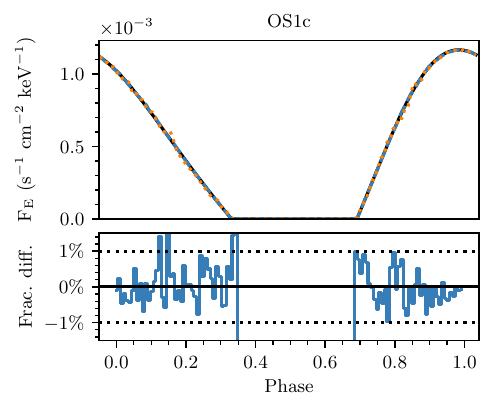}
\end{subfigure}
\begin{subfigure}{\os1width\textwidth}
 \includegraphics{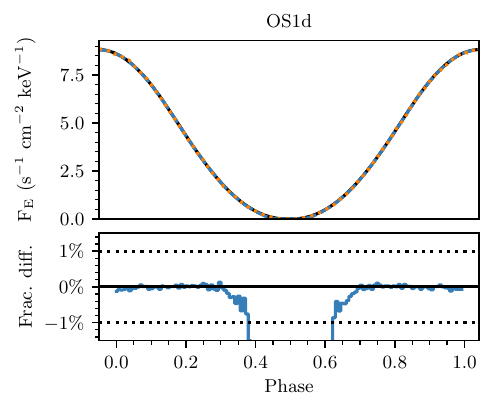}
\end{subfigure}
\begin{subfigure}{\os1width\textwidth}
 \includegraphics{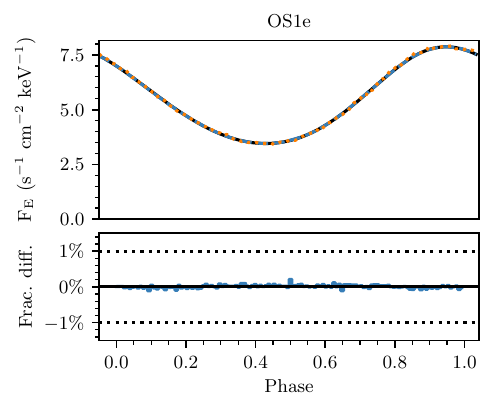}
\end{subfigure}
\begin{subfigure}{\os1width\textwidth}
 \includegraphics{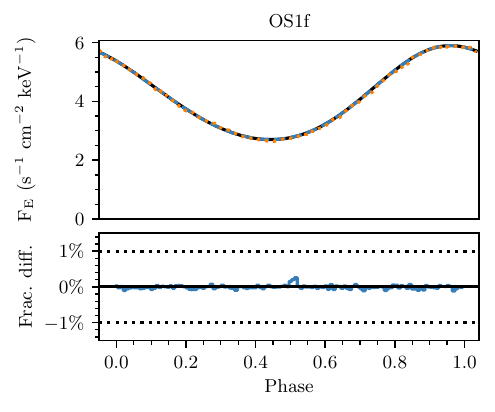}
\end{subfigure}
\caption{A comparison of the pulse profile predictions between \codename{} (dashed blue and dotted orange for the o2e and mc schemes, respectively) and \citetalias{bogdanov_constraining_2019} (black), for tests OS1a--OS1f. The parameters can be found in Table~\ref{tab:sd1}.\label{fig:os1_p1}}
\end{figure*}

\begin{figure*}
\centering
\begin{subfigure}{\os1width\textwidth}
 \includegraphics{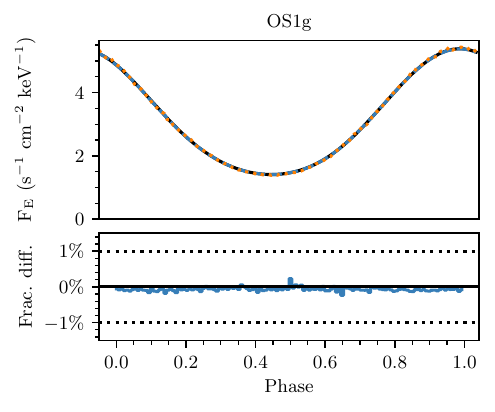}
\end{subfigure}
\begin{subfigure}{\os1width\textwidth}
 \includegraphics{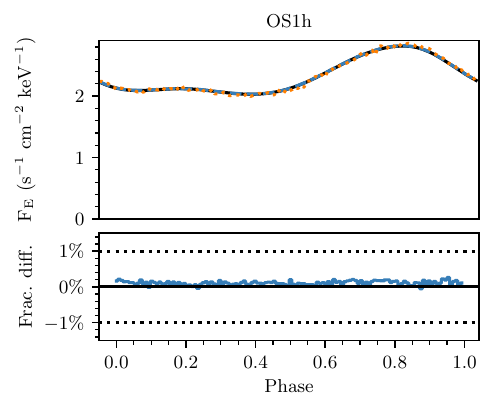}
\end{subfigure}
\begin{subfigure}{\os1width\textwidth}
 \includegraphics{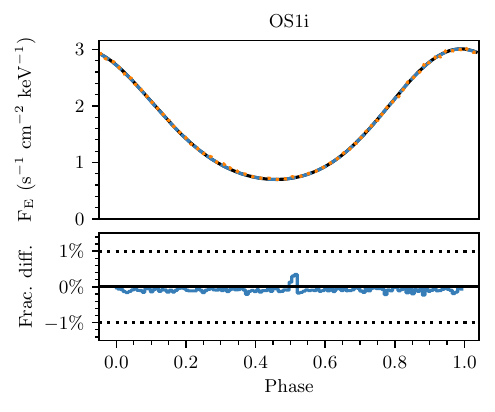}
\end{subfigure}
\begin{subfigure}{\os1width\textwidth}
 \includegraphics{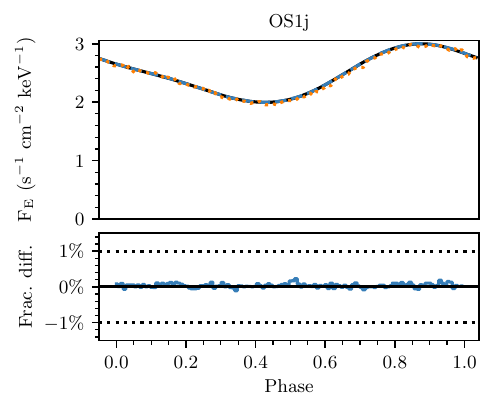}
\end{subfigure}
\caption{A comparison of the pulse profile predictions between \codename{} (dashed blue and dotted orange for the o2e and mc schemes, respectively) and \citetalias{bogdanov_constraining_2019} (black), for tests OS1g--OS1j. The parameters can be found in Table~\ref{tab:sd1}.\label{fig:os1_p2}}
\end{figure*}

\citetalias{bogdanov_constraining_2019} also computed 10 pulse profiles under the oblate Schwarzschild (OS) assumption, which are denoted OS1a--OS1j. Under the OS assumption, the exterior spacetime is still Schwarzschild, while the shape of the neutron star becomes oblate. Similar to the SD tests, a single circular hotspot is assumed. We verify \codename{} by computing the pulse profiles with identical parameters to those in \citetalias{bogdanov_constraining_2019} (tabulated in Table~\ref{tab:sd1}), assuming also an oblate shape for the neutron star surface. The pulse profiles computed by the two schemes of \codename{}, as well as the fractional difference with respect to \citetalias{bogdanov_constraining_2019}, are presented in Figs.~\ref{fig:os1_p1}--\ref{fig:os1_p2}. Again, the o2e results agree with \citetalias{bogdanov_constraining_2019} within 1\%, while the mc results have larger fluctuations.

\subsubsection{Oblate NS with hotspots of complex geometries}
\label{sec:com_geo}
\citet{choudhury_exploring_2024} (hereafter \citetalias{choudhury_exploring_2024}) performed several tests to compute the pulse profiles assuming more complex hotspot geometries, such as ring or crescent shapes. In addition to the hotspot geometry, the tests in \citetalias{choudhury_exploring_2024} differ from those in \citetalias{bogdanov_constraining_2019} in several other aspects. The specific intensity of surface radiation in \citetalias{choudhury_exploring_2024} is read from a look-up table generated by the \textsc{NSX} model \citep{ho_atmospheres_2001}, instead of being a simple blackbody spectrum. \citetalias{choudhury_exploring_2024} also convolved the phase-resolved flux density with the instrument responses of \textit{NICER} and took the interstellar attenuation into account.

\begin{figure*}
\begin{subfigure}{\c24width\textwidth}
 \includegraphics{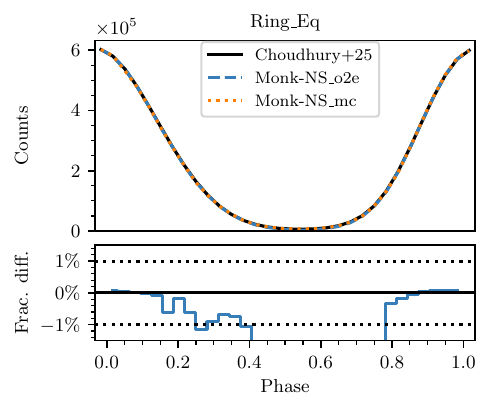}
\end{subfigure}
\begin{subfigure}{\c24width\textwidth}
 \includegraphics{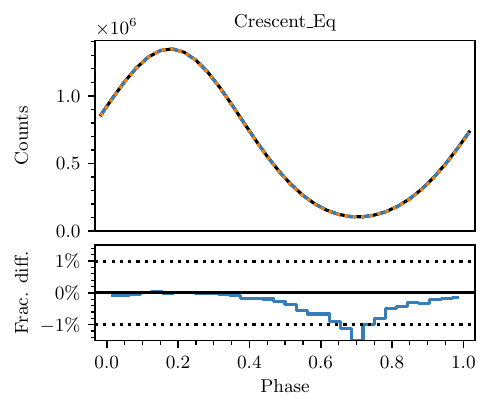}
\end{subfigure}
\begin{subfigure}{\c24width\textwidth}
 \includegraphics{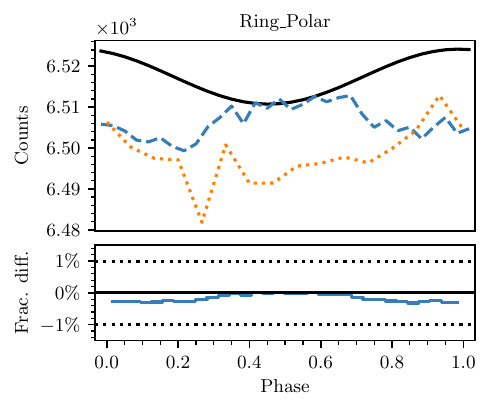}
\end{subfigure}
\begin{subfigure}{\c24width\textwidth}
 \includegraphics{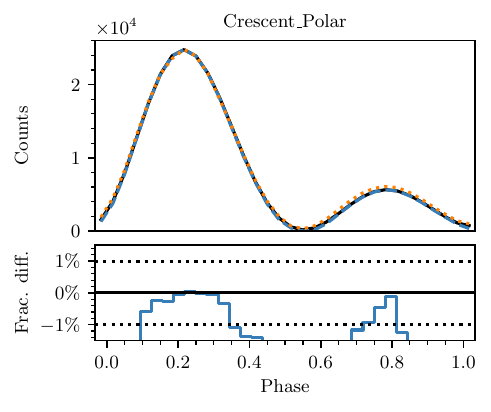}
\end{subfigure}
\caption{Comparison of the pulse profiles computed by \codename{} (dashed blue and dotted orange for the o2e and mc schemes, respectively) with \citetalias{choudhury_exploring_2024} (black), assuming oblate neutron stars with hotspots of complex geometries (see Table~\ref{tab:c24} for parameters).\label{fig:c24}}
\end{figure*}

For comparison, we compute pulse profiles using \codename{} with identical parameters that are tabulated in Table~\ref{tab:c24}. In our computations, we assume oblate NSs and use relevant files (NSX look-up table, NICER response files, and interstellar absorption cross-sections) provided by  \citetalias{choudhury_exploring_2024}\footnote{https://zenodo.org/records/13133749}. The pulse profiles computed with the two schemes, as well as the fractional differences compared to \citetalias{choudhury_exploring_2024}, are presented in Fig.~\ref{fig:c24}. Again, consistent results are obtained.

\begin{table*}
\begin{center}
\caption{The values of parameters for tests in Sec.~\ref{sec:com_geo}.\label{tab:c24}}
\begin{tabular}{lllllll}
\hline
Test & \multicolumn{3}{l}{hotspot} & \multicolumn{3}{l}{masking component} \\
& $\theta_0$ & $R_{\rm 0}$ & $\phi_0$ & $\theta_m$ & $R_{\rm m}$ & $\phi_m$ \\
& (rad) & (rad) & (rad) & (rad) & (rad) & (rad) \\
\hline
Ring-Eq & $\pi/2$ & 0.5 & $-0.2$ & $\pi/2$ & $0.25$ & 0 \\
Ring-Polar & $\pi - 0.001$ & 0.05 & $-0.02$ & $\pi - 0.001$ & 0.025 & 0\\
Crescent-Eq & $1$ & $1.4$ & $-1$ & 0.5 & $\pi/2 -0.001$ & 0 \\
Ring-Polar & 1.15 & 1.3 & $-0.1$ & 1.75 & $\pi/2 - 0.001$ & 0\\
\hline
\multicolumn{7}{l}{$\theta_0$: the centroid colatitude of the hotspot;} \\
\multicolumn{7}{l}{$R_{0}$: the angular radius of the hotspot;} \\
\multicolumn{7}{l}{$\phi_0$: the centroid azimuthal angle of the hotspot;}\\
\multicolumn{7}{l}{$\theta_m$, $R_m$, $\phi_m$: the centroid colatitude, angular radius, and centroid azimuthal angle of the mask.} \\
\end{tabular}
\end{center}
\end{table*}

\subsection{Photon redshift}
To validate the treatment of the photon redshift in \codename{}, we repeat the tests SD2a--SD2d and OS1k--OS1l in \citetalias{bogdanov_constraining_2019}, the parameters of which are tabulated in Table~\ref{tab:sd1}.
\footnote{We note a few typos in Table\,5 of \citetalias{bogdanov_constraining_2019} regarding the parameters of tests OS1k and OS1l. By comparing their Fig.\,14 with their Fig.\,8, it is easy to see that both tests actually assumed a spin frequency of 400\,Hz instead of 600\,Hz, and the angular radii are 0.01 and 1\,rad (instead of 1 and 0.01\, rad) for OS1k and OS1l, respectively.}

For each test, \citetalias{bogdanov_constraining_2019} computes the energy spectrum at various rotation phases of MSPs, while the surface emission is assumed to originate from a circular hotspot. In particular, in these tests, the rest frame emission of the hotspot is limited to a very narrow energy range around 1 keV (0.99998--1.00002\,keV and 0.995--1.005\,keV for NSs with spin periods of 1 and 400 Hz, respectively). With the o2e scheme, we take identical assumptions with \citetalias{bogdanov_constraining_2019}. While for the mc scheme, we simply assume the surface radiation to be mono-energetic with a co-moving frame energy of 1\,keV.

\begin{figure}
 \includegraphics[width=\columnwidth]{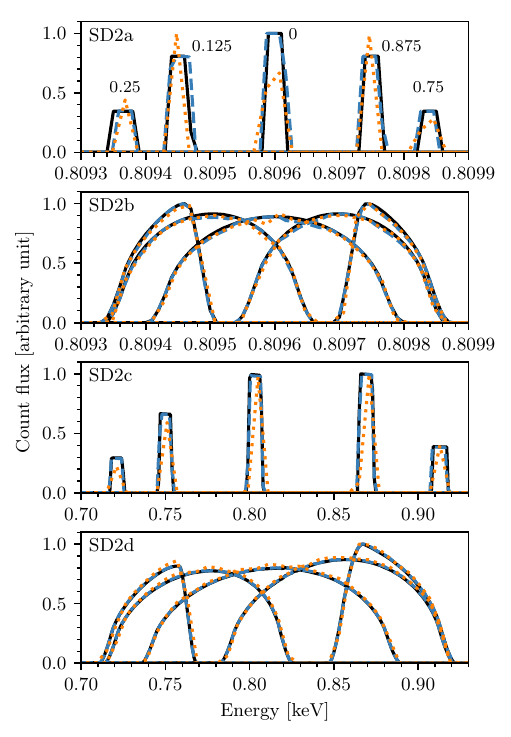}
 \caption{A validation on the computation of photon redshift in \codename{}. The energy spectra measured by the distant observer at various phases: 0.25, 0.125, 0, 0.875, 0.75, in the sequence of ascending observed line energy (also indicated in the top panel). In each panel, solid black, dashed blue, and dotted orange curves represent results of \citetalias{bogdanov_constraining_2019}, the o2e scheme of \codename{}, and the mc scheme of \codename{}, respectively. The parameters used in tests SD2a-SD2d can be found in Table~\ref{tab:sd1}.\label{fig:redshift}}
\end{figure}

\begin{figure}
 \includegraphics[width=\columnwidth]{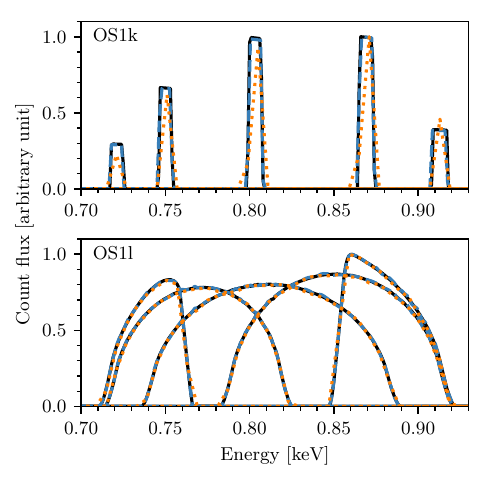}
 \caption{A validation on the photon redshift of \codename{}. The same with Fig.~\ref{fig:redshift}, but for OS tests. The parameters used in tests OS1k and OS1l can be found in Table~\ref{tab:sd1}.\label{fig:redshift_os}}
\end{figure}

The SD and OS results are presented in Figs.~\ref{fig:redshift} and \ref{fig:redshift_os}, respectively. The o2e scheme agrees with \citetalias{bogdanov_constraining_2019} quite well, in both the energy shift and the spectral shape. The mc scheme agrees well with \citetalias{bogdanov_constraining_2019} for cases with hotspots of 1 rad in radius, and obtained different spectral shapes for cases with hotspots of 0.01\,rad (SD2a, SD2c, and OS1k) but still produces identical energy shifts. The difference in the spectral shape between the mc scheme and \citetalias{bogdanov_constraining_2019} is simply due to different assumptions on the local spectrum, as mentioned earlier. The agreement in the energy shift demonstrates that \codename{} correctly computes the redshift.

\subsection{Polarisation of X-ray bursts}
We compute the X-ray polarisation of neutron stars during burst oscillations as a function of the rotational phase. We take the assumption of \citet{viironen_light_2004} that during the thermonuclear burst, the X-ray emission is emitted by a hotspot on the surface of the neutron star. The photons produced in the deep layer of the neutron star atmosphere experience multiple scatterings before finally escaping the surface, and the polarisation and limb-darkening properties of the emergent radiation follow the Chandrasekhar-Sobolev solutions for a semi-infinite plane \citep{chandrasekhar_radiative_1960,sobolev_treatise_1963}. Specifically, in the rest frame, the polarisation of the escaping radiation is perpendicular to the meridian plane, and the polarisation degree is a function of $\mu\equiv {\rm cos}{\vartheta}$:
\begin{equation}
\label{eq:chandra_delta}
\delta = 0.117 \frac{1-\mu}{1+3.582\mu},
\end{equation}
and the specific intensity
\begin{equation}
\label{eq:chandra_flux}
 I_\mu \propto 1 + 2.08\mu.
\end{equation}
These two approximate expressions follow \citet{viironen_light_2004}.

\begin{figure*}
\begin{subfigure}{\burstwidth\textwidth}
 \includegraphics{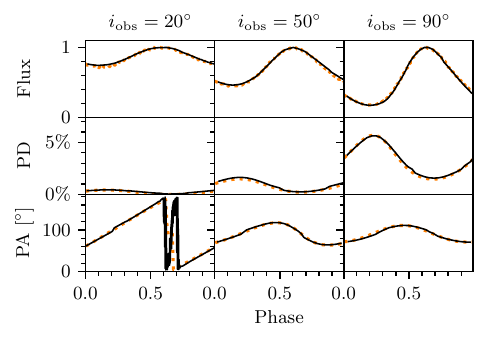}
 \end{subfigure}
\begin{subfigure}{\burstwidth\textwidth}
 \includegraphics{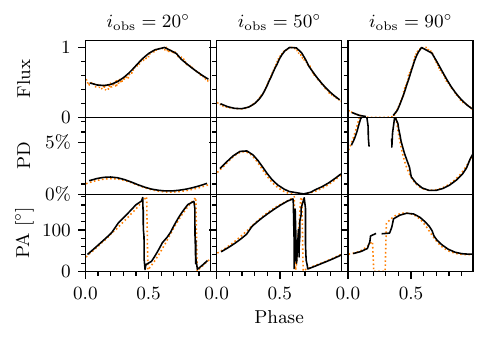}
 \end{subfigure}
\caption{A comparison of the model predictions of the X-ray polarisation of thermonuclear bursts between \codename{} (black) and \textsc{ARCMANCER} (orange). The parameters are: $M_{\rm NS}=1.6~\rm M_\odot$, $R_{\rm eq}=12~\rm km$, $\nu_{\rm spin}=700~\rm Hz$. The hotspot is located at a colatitude of $20^\circ$ and $50^\circ$, respectively, in the left and right panels. In each panel, we compare the pulse profile (top), polarisation degree (middle), and polarisation angle(bottom). We also present results of three different inclinations: $20^\circ$, $50^\circ$, and $90^\circ$, as indicated in the plot.\label{fig:burst_pol}}
\end{figure*}

We compare the results with \citet{pihajoki_general_2018}\footnote{Since we could not find a data file associated with \citetalias{pihajoki_general_2018}, we extracted data directly from the figures in the paper using WebPlotDigitizer (http://arohatgi.info/WebPlotDigitizer/), an online tool that digitizes data from plots. This process provided us with only approximate values of their results, preventing us from conducting a detailed quantitative comparison between the results.}(hereafter \citetalias{pihajoki_general_2018}), using identical parameters: the neutron star has $M_{\rm NS}=1.6~\rm M_\odot$, $R_{\rm eq}=12~\rm km$, and $\nu_{\rm spin} = 700~\rm Hz$. The pulse profiles, polarisation degree (PD), and polarisation angle (PA)\footnote{To compare with \citetalias{pihajoki_general_2018}, we add $\pi/2$ to their polarisation angle due to different conventions used in \codename{} and \textsc{ARCMANCER}.} for two different colatitude of the hotspot are presented in the left and right panels of Fig.~\ref{fig:burst_pol}, respectively. Consistency between \codename{} and \citetalias{pihajoki_general_2018} is evident.

\subsection{Convergence of the two numerical schemes}
\begin{figure}
 \includegraphics[width=\columnwidth]{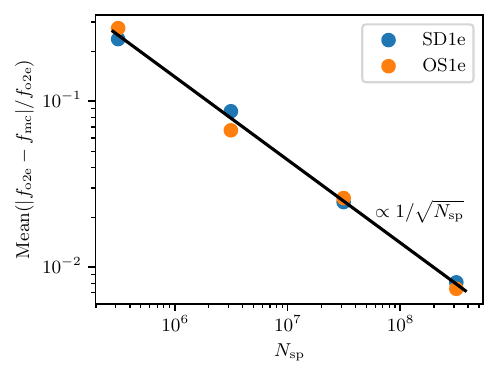}
 \caption{Scalability of the mc scheme. The mean fractional difference between the mc and o2e schemes as a function of the $N_{\rm sp}$, the number of superphotons sampled in the mc scheme, are plotted in solid circles. The results for test cases SD1e and OS1e, are plotted in blue and orange colors, respectively. We also plot a $\propto 1/\sqrt{N_{\rm sp}}$ curve for comparison.\label{fig:scaling}}
\end{figure}

In this subsection, we demonstrate that the results of the mc scheme converge to the o2e scheme as the number of superphotons increases. For this purpose, we compute the pulse profiles of two cases: SD1e and OS1e, using the mc scheme, and for each case, we consider several different numbers of superphotons. We then examine the fractional difference between the mc and the more accurate o2e results. The mean fractional difference between the two schemes as a function of the number of superphotons in the mc scheme is presented in Fig.~\ref{fig:scaling}. For both cases, the fractional difference between the schemes scales with $1/\sqrt{N_{\rm sp}}$, as expected for a Monte Carlo scheme.

\section{Application: low pulsation amplitude NSLMXBs}
\label{sec:application}

We apply \codename{} to study the pulsation amplitude of NSLMXBs. It remains a puzzle why pulsations have not been detected in most NSLMXBs. Several models have been proposed, including: the magnetic field screening \citep[e.g.][]{cumming_magnetic_2001}, the scattering \citep[e.g.][]{brainerd_effect_1987,kylafis_effects_1987,wang_smearing_1987, bussard_effect_1988,titarchuk_why_2002}, the alignment of the magnetic and spin axes \citep{lamb_origin_2009, lamb_model_2009}, and gravitational light-bending \citep{wood_effect_1988,ozel_what_2009}. In this section, we investigate the X-ray polarisation properties of these scenarios to see if X-ray polarimetric observations can differentiate the proposed models.

\subsection{The magnetic field screening model}
\label{sec:magnetic_screen}
In this model, the NS magnetic field is buried by the unmagnetised material that is accreted onto the neutron star surface \citep[e.g.][]{cumming_magnetic_2001}. The X-ray polarimetric properties of such low-magnetized accreting neutron stars have been studied by \citet{gnarini_polarization_2022}, who found that, in general, the PD is sensitive to the observer's inclination.

\subsection{The scattering model}
\label{sec:scattering}
In this model, the pulsed emission from the NS surface is reprocessed by a Comptonizing cloud before reaching the observer, thus the pulsation amplitude is reduced \citep[e.g.][]{brainerd_effect_1987,kylafis_effects_1987,wang_smearing_1987, bussard_effect_1988,titarchuk_why_2002}. In this work, we investigate this model using the mc scheme of \codename{}, as this cannot be handled by the o2e method. We consider a spherical corona that surrounds the surface of the neutron star, which may physically mimic, for example, the transition layer between the neutron star and the outer accretion disc. We assume a NS with mass of $1.4~\rm M_\odot$ and radius of $12~\rm km$, and a circular hotspot at a colatitude of $90^\circ$ and with a radius of $10^\circ$.
The values of other parameters are taken from \citet{titarchuk_why_2002}: the temperature of the hotspot $T_{\rm NS}=0.69~\rm keV$, the coronal temperature $kT_e=19.5~\rm keV$, and the outer radius of the corona $R_{\rm c, out}=60~\rm km$. We experiment with different values of the Thomson optical depth of the corona $\tau_{\rm T} =1, 2, 4, 6$ (defined as $n_e \sigma_{\rm T} (R_{\rm c,out}-R_{\rm eq})$ with $n_e$, $\sigma_{\rm T}$ being the electron density and the Thomson scattering cross section, respectively) to examine the effect of the scattering on the emergent lightcurve.

\begin{figure*}
 \includegraphics[width=\textwidth]{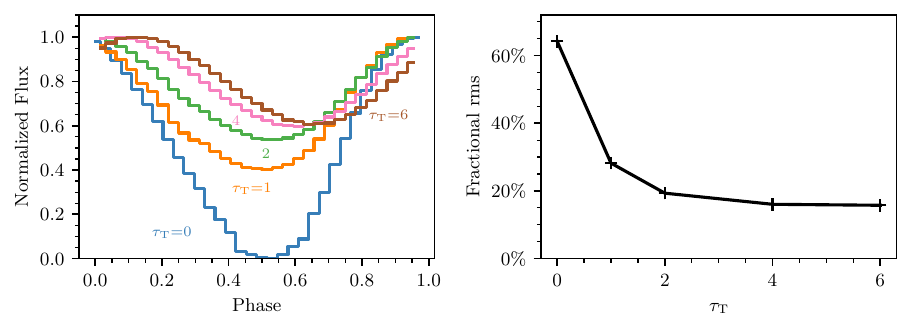}
 \caption{Left: the pulse profiles of rotating neutron stars that are reprocessed by spherical coronae. The surface radiation of the neutron stars is assumed to originate from a circular hotspot. Different colors indicate different Thomson optical depths of the coronae. Right: the fractional rms of the phase-resolved flux as a function of the optical depth. 
 \label{fig:corona_sphere}}
\end{figure*}

The pulse profiles corresponding to different optical depths are presented in the left panel of Fig.~\ref{fig:corona_sphere}. The amplitude of the pulsation decreases with the optical depth, which is visible in the pulse profiles, and quantitatively confirmed by examining the variation of the fractional rms of the phase-resolved flux with $\tau_{\rm T}$ (middle panel of Fig.~\ref{fig:corona_sphere}). The fractional rms drops rapidly with increasing $\tau_{\rm T}$ when $\tau_{\rm T} \lesssim 2$, and stays around $20\%$ for $\tau_{\rm T}=4$ and $\tau_{\rm T}=6$. The fractional rms only drops by a factor of three as $\tau_{\rm T}$ increases from $0$ to $6$, which appears inconsistent with the analytical result of \citet{titarchuk_why_2002}, who claimed that the rms is expected to decrease rapidly when $\tau_{\rm T} \geq 4$ using the parameters adopted in this work. The discrepancy likely arises from differing underlying assumptions. In \citet{titarchuk_why_2002}, the Green’s function is computed for an isotropic seed source with sinusoidal variability located at the center of a scattering cloud. In contrast, our work employs Monte Carlo simulations to evaluate the response for a beamed seed source, which has been shown to exhibit different behavior \citep[e.g.,][]{kylafis_effects_1987}. This discrepancy underscores the necessity of performing Monte Carlo simulations. While a comprehensive exploration of the parameter space is needed to fully understand this scenario, this is beyond the scope of this work. Nevertheless, we investigate the polarisation properties for the case of a large optical depth of $\tau_{\rm T}=6$. The 2--8\,keV\footnote{The effective energy bands of \textit{IXPE} and \textit{eXTP}/PFA.} polarisation degree as measured by observers at different inclinations is found to be extremely low ($< 0.5\%$) for all inclinations.

\subsection{The axes-alignment model}
\label{sec:alignment}
In this model, the magnetic poles are pushed toward the spin axis due to accretion \citep{lamb_origin_2009, lamb_model_2009}, and are nearly perfectly aligned with the spin axis in the LMXB phase. Under such a symmetric configuration, the pulsation amplitude remains small at any inclination.

\begin{figure}
 \includegraphics[width=\columnwidth]{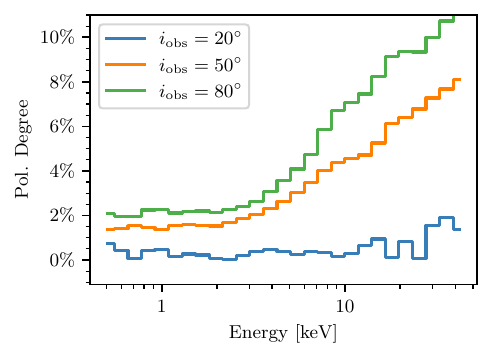}
 \caption{The polarisation degree of hotspot emission in relation to photon energy from neutron stars with aligned spin and magnetic axes in LMXBs. The results corresponding to various inclinations are plotted in different colors, as indicated in the plot.\label{fig:pol_aligned}}
\end{figure}

Using \codename{}, we compute the X-ray polarisation of a neutron star with aligned spin and magnetic axes. The X-ray emission is assumed to be emitted by two antipodal hotspots centered at the north and south poles of the neutron star, respectively. Following \citet{ozel_what_2009}, we take the angular radius of the hotspots to be $20^\circ$. We compute the local emission of the hotspots by assuming a plane-parallel atmosphere with the following parameters: electron temperature $kT_e=50~\rm keV$, Thomson optical depth $\tau_{\rm T}=1$, and seed photon temperature $kT_{\rm bb} = 1~\rm keV$, following \citet{salmi_neutron_2021}. The polarisation degree as seen by observers at various inclinations are presented in Fig.~\ref{fig:pol_aligned}, while the polarisation angle is aligned with the rotation axis of the neutron star for all inclinations and photon energies. A strong dependence of the PD on the observer's inclination is observed.

\subsection{The gravitational light-bending model}
\label{sec:light_bending}
This model proposes that LMXBs host more massive neutron stars than AMXPs, therefore experiencing stronger gravitational effects, given that the neutron star radius remains approximately constant with mass for most EoSs. The general relativistic effects, particularly the light-bending effect, lead to a decrease in the pulsation amplitude.

\begin{figure*}
\centering
\includegraphics[width=\textwidth]{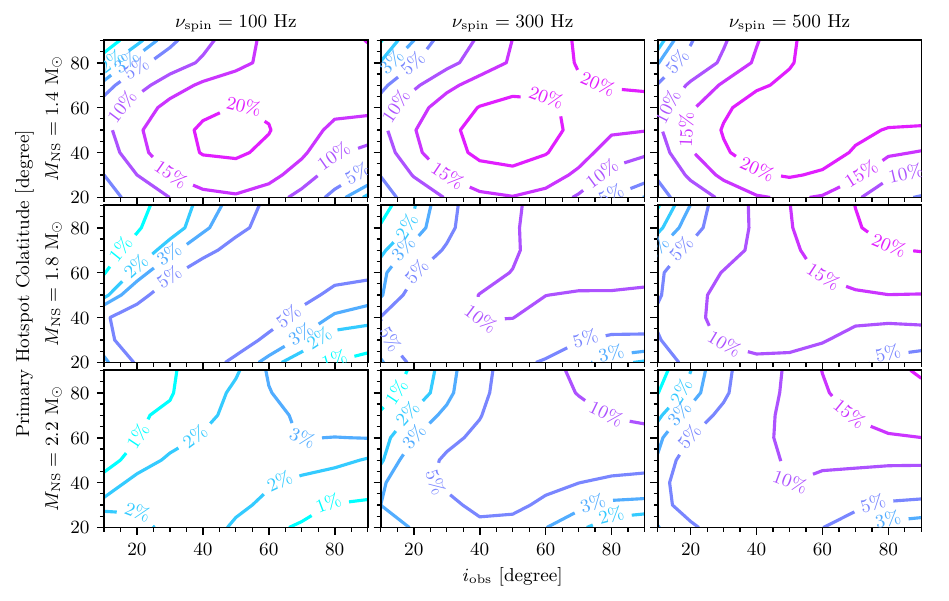}
\caption{The fractional rms of the phase-resolved flux of AMXPs with mass of $1.4$ (upper panels), $1.8~\rm M_\odot$ (middle panels), and $2.2~\rm M_\odot$ (bottom panels) as a function of the colatitude of the primary hotspot and the observer's inclination. Here we assume that the emission originates from two antipodal hotspots on the surface of the neutron star. We present results for NSs of different spin frequencies: 100, 300, and $500~\rm Hz$, from left to right. \label{fig:rms}}
\end{figure*}

To investigate this model, we compute with \codename{} the pulse profiles of massive neutron stars with $M_{\rm NS}=1.8$ and $2.2~\rm M_\odot$ and a radius of 12\,km. We assume that two antipodal hotspots exist on the neutron star surface and take the radius of the hotspots to be $20^\circ$, following \citet{ozel_what_2009}. In Fig.~\ref{fig:rms}, we show the fractional rms of the phase-resolved flux as a function of the colatitude of the primary hotspot $\theta_0$ and the observer's inclination $i_{\rm obs}$ for three different spin frequencies: $\nu_{\rm spin}=100, 300, 500~\rm Hz$, presented in the left, middle, and right panels, respectively. The results for $M_{\rm NS}=1.8$ and $2.2~\rm M_\odot$ are plotted in the middle and lower panels, respectively. For comparison, we also present the results for $M_{\rm NS}=1.4~\rm M_\odot$ in the upper panels. Our results confirm the conclusions of \citet{wood_effect_1988} and \citet{ozel_what_2009} that the pulsation amplitude drops dramatically as the neutron star mass increases. For the case of $M_{\rm NS}=2.2~\rm M_\odot$ and $\nu_{\rm spin}=100~\rm Hz$, the rms is below 5\% for all combinations of $\theta_0$ and $i_{\rm obs}$. In certain regions of the $\theta_0-i_{\rm obs}$ plane, namely low-$\theta_0$/high-$i_{\rm obs}$ or high-$\theta_0$/low-$i_{\rm obs}$, the rms amplitude is even smaller than $\sim$2\%. However, as the spin frequency increases (middle and right panels of Fig.~\ref{fig:rms}), the pulsation amplitude rises due to the enhanced Doppler beaming effect \citep[e.g.][]{viironen_light_2004}. This suggests that for the light-bending model to be effective, the neutron star spin frequency cannot be too high.

\begin{figure}
 \includegraphics[width=\columnwidth]{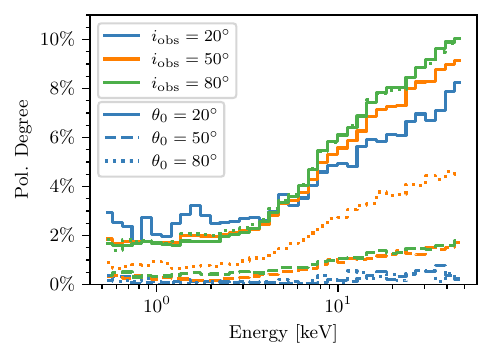}
 \caption{The phase-averaged polarisation degree of the emission from two antipodal hotspots on the surface of neutron stars with a mass of $1.8~\rm M_\odot$. The results corresponding to various inclinations are plotted in different colors, as indicated in the plot. Solid, dashed, and dotted curves represent cases with primary hotspot colatitude of $20^\circ$, $50^\circ$, and $80^\circ$, respectively.\label{fig:highmass}}
\end{figure}

We proceed to compute the polarisation of high-mass AMXPs with $M_{\rm NS}=1.8~\rm M_\odot$ and $\nu_{\rm spin}=100~\rm Hz$. We assume two antipodal hotspots, each with angular radius of $20^\circ$. The local emission of the hotspot is assumed to be emerging from a plane-parallel atmosphere with $kT_{\rm e}=50~\rm keV$, $\tau_{\rm T}=1$, and $kT_{\rm bb}=1~\rm keV$. The PD is presented in Fig.~\ref{fig:highmass}, while the PA is either parallel or perpendicular to the spin axis and is constant with energy. For a fixed hotspot colatitude, the PD increases with the observer's inclination, a behavior that is the same with the aligned axes scenario. However, for high-mass AMXPs, the PD is also sensitive to the colatitude of the hotspot, as one can see by looking at curves of different line styles in Fig.~\ref{fig:highmass}.

\subsection{Using X-ray polarisation to differentiate the models}
\begin{figure}
\centering
\includegraphics[width=\columnwidth]{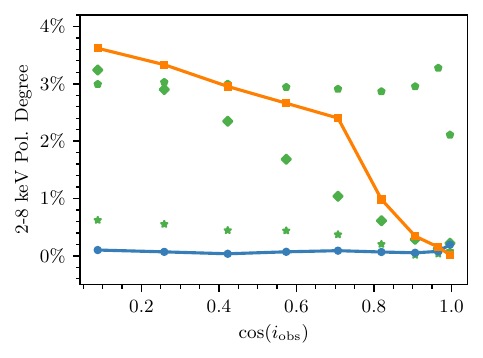}
\caption{The dependence of the 2--8\,keV PD on the observer's inclination, as predicted by different models for the low pulsation amplitude of NSLMXBs. Blue: the scattering model (see Sec.~\ref{sec:scattering} for full parameters), with the Thomson optical depth $\tau_{\rm T}=6$. Orange: the axes-alignment model (see Sec.~\ref{sec:alignment} for full parameters). Green: the light-bending model (see Sec.~\ref{sec:light_bending} for full parameters) with $M_{\rm NS}=1.8~\rm M_\odot$. Pentagon, diamond, and star shapes represent results obtained assuming the primary hotspot's colatitude of $20^\circ$, $50^\circ$, and $80^\circ$, respectively.\label{fig:all}}
    
\end{figure}

As outlined in Sec.~\ref{sec:magnetic_screen}--\ref{sec:light_bending}, for different models explaining the low pulsation amplitude of NSLMXBs, we expect distinct inclination-dependent behavior of of the PD. A direct comparison of the dependence of the 2--8\,keV PD on the inclination, as computed by \codename{} assuming various scenarios, is presented in Fig.~\ref{fig:all}. In the magnetic screening or axis-alignment model, the PD is sensitive to the inclination. The scenario where the neutron star surface radiation is screened by an optically-thick spherical corona predicts a consistently low PD regardless of the inclination. If the low pulsation amplitude is due to light-bending, the PD is determined jointly by both the observer's inclination and the magnetic angle (the angle between the spin and magnetic axes). Since evolution of the magnetic angle depends on the accretion history, and this model does not predict a fixed magnetic angle across different NSLMXBs, we anticipate the PD to be insensitive to the observer's inclination if we examine a sample of NSLMXBs.
In summary, these models can be differentiated by characterizing the inclination dependence of PD across a sample of NSLMXBs.

\section{Application: polarisation properties of hotspots of complex geometries}
\label{sec:com_geo_pol}

In this section, we apply \codename{} to investigate the effect of hotspot shape on the radiation properties as seen by observers. While most of previous studies assumed circular hotspots (\citetalias{choudhury_exploring_2024} as an exception), the actual hotspots may have more complex morphologies. For instance, two oval rather than circular hotspots are required to explain the pulse profiles of PSR J0030+0451, one of which is even highly elongated \citep{miller_psr_2019}. It is therefore essential to examine if the radiation properties obtained under the assumptions of circular hotspots change significantly when considering hotspots with more complex shapes. In this study, we consider two different shapes. The first complex hotspot shape we investigate is an elongated ellipse.
The second hotspot shape is a crescent shape that mimics two overlapping circular hotspots.


To achieve our goal, we compute the flux, polarisation degree, and polarisation angle of rotating neutron stars with hotspots of complex shapes and compare the results with circular hotspots. For all hotspot morphologies, we assume a neutron star of $M_{\rm NS}=1.4~\rm M_\odot$ and $R_{\rm eq} = 12~\rm km$, and assume that the emission originates from two identical, antipodal hotspots with temperature of $0.35~\rm keV$. We further assume that, in the rest frame of the hotspot, the angular distribution of both the specific intensity and the polarisation degree follows the Chandrasekhar formula (Eqs.~\ref{eq:chandra_delta}--\ref{eq:chandra_flux}) that are appropriate for a semi-infinite scattering atmosphere.

The elliptical hotspot is assumed to be highly elongated with a minor-to-major ratio of 0.2, whose major axis is aligned with a curve of constant colatitude. The crescent hotspot is constructed by obscuring a circular hotspot with a circular mask of the same centroid colatitude and angular size, offset by $10^\circ$ in azimuth. We experiment with different parameters: neutron star spin frequencies of $1$ and $700~\rm Hz$, hotspot angular size of $20^\circ$ and $40^\circ$ \footnote{For highly elongated hotspots, the angular size is defined with respect to the major axis.}, and hotspot centroid colatitude of $30^\circ$ and $90^\circ$.

\begin{figure*}
\begin{subfigure}{\burstwidth\textwidth}
 \includegraphics{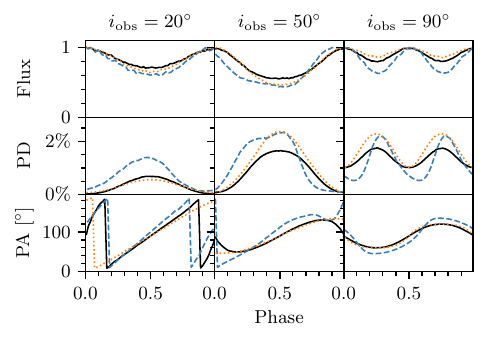}
 \end{subfigure}
\begin{subfigure}{\burstwidth\textwidth}
 \includegraphics{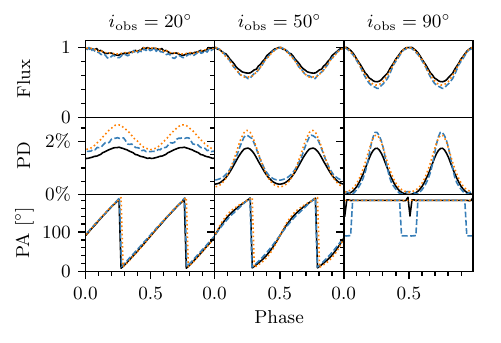}
 \end{subfigure}
\caption{A comparison of the normalized flux (top), polarisation degree (middle), and polarisation angle (bottom) of slowly rotating neutron stars ($\nu_{\rm spin}=1~\rm Hz$) with hotspots of complex geometries. The results of ellipse and crescent hotspots are plotted in blue dashed and orange dotted lines, respectively, whereas the results of circular hotspots are presented in solid black lines for comparison. The left and right subfigures present the results of hotspots centered at colatitude of $30^\circ$ and $90^\circ$, respectively.\label{fig:irr_onehz}}
\end{figure*}

\begin{figure*}
\begin{subfigure}{\burstwidth\textwidth}
 \includegraphics[width=\columnwidth]{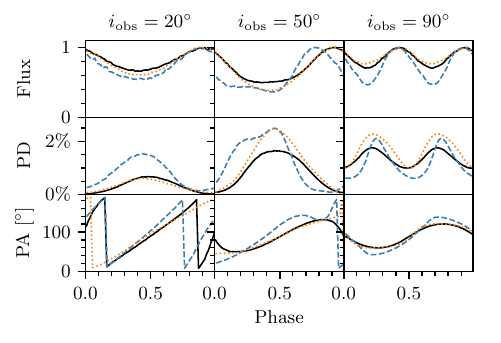}
 \end{subfigure}
\begin{subfigure}{\burstwidth\textwidth}
 \includegraphics[width=\columnwidth]{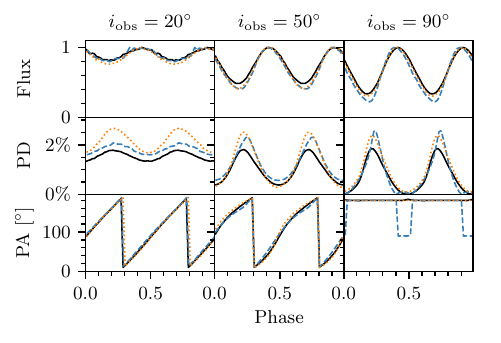}
 \end{subfigure}
\caption{The same with Fig.~\ref{fig:irr_onehz}, but for rapidly rotating neutron stars ($\nu_{\rm spin}=700~\rm Hz$).\label{fig:irr_hz700}}
\end{figure*}

The results for spin frequencies of $1$ and $700~\rm Hz$ are presented in Figs.~\ref{fig:irr_onehz} \& \ref{fig:irr_hz700}, respectively. We find the results to be insensitive to the angular size of the hotspot, therefore, for brevity, we present results only for a hotspot angular size of $40^\circ$. For both slowly and rapidly rotating neutron stars, the polarisation degree is sensitive to the shape of the hotspot, with larger PD contributed by complex shapes compared with circular hotspots, even if the pulse profiles are identical, e.g.\ for $i_{\rm obs}=20^\circ$ and $\theta_0=30^\circ$ (the left panels of the subfigure to the right, for Figs.~\ref{fig:irr_onehz} \& \ref{fig:irr_hz700}). On the other hand, the profiles of PD and PA are rather insensitive to the spin frequency.



\section{Summary}
\label{sec:summary}
We customize the general relativistic radiative transfer code \monk{} to model the surface radiation of rotating neutron stars and developed \codename{}, which assumes the Kerr or the Hartle-Thorne spacetime exterior to the surface of the neutron star. Besides assuming a spherical star, \codename{} can also handle an oblate star configuration. We implement two different numerical schemes: the o2e scheme, in which we set up an image plane at the observer and generate a synthetic image by raytracing photons backward in time from the observer to the neutron star surface, and the mc scheme, in which photons are randomly sampled at the neutron star's surface.

Compared to purely analytical treatments, \codename{} can handle more complex geometries and/or physical conditions, especially when the surface radiation is further reprocessed by an extended scattering medium. With \codename{}, we can simultaneously compute the energy spectrum, polarisation, and lightcurves of rotating neutron stars. This capability is particularly important for future X-ray missions that can obtain high-quality spectral, timing, and polarimetric measurements concurrently, for example, \textit{The Enhanced X-ray Timing and Polarimetry mission} (\textit{eXTP}; \citealt{zhang_enhanced_2019}).

We validate \codename{} by performing a series of benchmarking tests in which we computed the energy spectrum, pulse profile, and polarisation of rotating neutron stars, and compared the results with those in existing literature, such as the energy spectrum with \citetalias{bogdanov_constraining_2019}, the pulse profiles with \citetalias{bogdanov_constraining_2019} and \citetalias{choudhury_exploring_2024}, and the polarisation results with \citetalias{pihajoki_general_2018}. We find the results of both numerical schemes employed by \codename{} agree with those reported in the literature. Therefore, with \codename{}, we are able to obtain comparably accurate constraints on NS parameters with other numerical codes. In particular, the energy spectrum and the pulse profiles computed with the o2e scheme agree with \citetalias{bogdanov_constraining_2019} and \citetalias{choudhury_exploring_2024} within 1\% except for phase bins where the flux is negligible or equal to zero. The results of the mc scheme converge to the o2e scheme as the number of superphotons increases, with the fractional difference between the two schemes inversely proportional to the square root of the number of photons sampled, as expected for a Monte Carlo scheme.

We apply \codename{} to study various models in order to explain why most NSLMXBs do not exhibit pulsations (Sec.~\ref{sec:application}). These models include the magnetic field screening model, the scattering model, the axis-alignment model, and the light-bending model. We compute the X-ray polarisation for different scenarios and find that, under the light-bending model, the PD of NSLMXBs is less dependent or even nearly independent of the observer’s inclination, compared to the magnetic field screening or axis-alignment models. In the scenario in which the surface radiation gets scattered by a spherical corona covering the neutron star, the PD is expected to be extremely low regardless of the observer's inclination. These results suggest that X-ray polarimetric observations can be used to distinguish between these models.

We also investigate the radiation properties of neutron stars with complex hotspot morphologies (Sec.~\ref{sec:com_geo_pol}). We find the PD to be sensitive to the hotspot shape, even if the pulse profiles are identical across different hotspot morphologies.
The predictions we obtain in the two applications (Sec.~\ref{sec:application} \& \ref{sec:com_geo_pol}) can serve as examples for cross-checks and tests of other neutron star surface emission models.

\section*{acknowledgements}
We are grateful to the referee for their careful reading and insightful comments, which have substantially improved the quality and clarity of this work. We thank Devarshi Choudhury for kindly providing the data of their paper. We also thank Michal Dov\v{c}iak, Michal Bursa, Vladim\'ir Karas, and Martin Urbanec for useful discussion. WZ would like to acknowledge the support by the National Natural Science Foundation of China (grants 12573019, 12333004, 12433005), and the support by the Strategic Priority Research Program of the Chinese Academy of Sciences through the grant XDB0550201. WY would like to acknowledge the support by the National Natural Science Foundation of China (grant 12373050). 

\section*{data availability}
The authors agree to make simulation data supporting the results in this paper available upon reasonable request.

\bibliographystyle{mnras}
\bibliography{ns_pulse}

\label{lastpage}
\end{document}